\documentclass[10pt]{article}
\usepackage[dvips]{graphicx}
\begin{document}

\begin{center}
\textbf{Voltage dependence of Hodgkin-Huxley rate functions for a 
multi-stage $K^+$ channel voltage sensor within a membrane}
 \end{center}
\begin{center} S. R. Vaccaro  \end{center}
\begin{center} 
{\em Department of Physics, University of Adelaide, Adelaide, South Australia, 
5005, 
Australia} \\
  \end{center}

{\em E-mail address:}\\
 {\em svaccaro@physics.adelaide.edu.au}\\

{\bf Abstract}

\begin{quotation}
 The activation of a $K^+$ channel sensor in two sequential stages during
a voltage clamp may be described as the translocation of a Brownian particle
 in an energy landscape with two large barriers between states. A solution 
of the Smoluchowski equation for a square-well approximation to the potential
 function of the S4 voltage sensor satisfies a master equation, and  has two
frequencies that may be determined from the forward and backward rate
functions. When the higher frequency terms have small amplitude, the
solution reduces to  the relaxation of a rate equation, where the derived
two-state rate functions are dependent on the relative magnitude of the
forward rates ($\alpha$ and $\gamma$) and the backward rates ($\beta$
and $\delta$) for each stage. In particular, the voltage dependence of the
Hodgkin-Huxley rate functions for a $K^+$ channel may be derived by assuming
that the rate functions of the first stage are large relative to those of
the second stage - $\alpha \gg \gamma $ and $\beta \gg \delta$. 
For a {\em Shaker} IR $K^+$ channel, the first forward and backward 
transitions  are rate  limiting ($\alpha < \gamma $ and $\delta \ll \beta$), and for an 
activation process with either two or three stages, the derived
two-state rate functions also have a voltage dependence that is of  a
similar form to that determined  for the squid axon.
The potential variation generated by the interaction between a two-stage
 $K^+$ ion channel and a noninactivating  $Na^+$  ion channel  is determined by the master
 equation for $K^+$ ion channel activation and the ionic current equation when 
the $Na^+$  ion channel activation time is small, and if $\beta \ll \delta $
 and $\alpha \ll \gamma $, the system may exhibit a small amplitude 
oscillation between spikes, or mixed-mode oscillation, in which the 
slow closed state modulates the $K^+$ ion channel conductance in the membrane.

\end{quotation}
\newpage

{\bf INTRODUCTION}

Based on the the measurement of increased $Na^+$  and $K^+$ ion channel conductance
during a depolarizing voltage clamp of the squid axon membrane, Hodgkin
and Huxley (HH) proposed a model of the action potential which  accounted
for the threshold potential, the refractory period and the speed of
transmission of the impulse \cite{hh}. The $Na^+$  and $K^+$ ionic
conductance was described by activation variables $m$ and $n$
and an inactivation variable $h,$ that each
satisfy a first-order rate equation with empirical transition rate
functions that depend on the potential difference $V$ across
the membrane \cite{hh}. It was assumed that the voltage sensitivity of $m$, $n$ and $h$
is dependent on the transverse movement of charged gating particles when the electric field
within the membrane is changed.
The existence of a delay in the increase of the $K^+$
and $Na^+$  conductance, and the transient nature of the $Na^+$  conductance increase
were represented by  $g_{K}\propto n^{4}$  and  $g_{Na}$ $\propto $ $m^{3}h$.
 The HH model has been applied to ion channels in
cardiac Purkinje fibres  \cite{nob}, toad myelinated neuron  \cite{fh},
 gastropod neuron (including an A-type inactivating
$K^+$ channel) \cite{cs}, and a bursting pacemaker neuron in the mollusc 
{\em Aplysia} \cite{pl}. 

The activation variable for the squid axon $K^+$ channel satisfies
a rate equation
\begin{equation}
\frac{dn}{dt}=\alpha_{n}-(\alpha_{n}+\beta_{n})n,  \label{raten}
\end{equation}
where the  empirical forward and backward rate functions are of the form 
\begin{eqnarray}
\alpha _{n} & = & \frac{C(V-A)}{1-\exp [-B(V-A)]}, \label{an} \\
\beta _{n}  & = & E\exp [-DV], \label{bn}
\end{eqnarray}
$V$ is the membrane voltage, and A to E are constants. The HH rate functions 
$\alpha_{n}$ and $\beta_{n}$ have been successful in describing activation in a wide variety of
ion channels because $\alpha_{n}$ represents the exponential dependence on 
$V$  for small depolarizations and the almost linear dependence on $V$ 
for large clamp potentials, and $\beta_{n}$  represents the exponential
voltage dependence of the rate constant for a large hyperpolarization.
The expression for $\alpha_{n}$ in Eq. (\ref{an}) may be obtained from a
 solution to the Smoluchowski equation for the probability density of states
of the voltage sensor, when the potential function is linear in the tranverse coordinate  $Z$ \cite{gh,va0}, and a rate equation for activation may be derived if there is
 a large diffusion or potential barrier between closed and open states \cite{va1}.
 However, in view of the presence of  negative residues on the S2 and S3 
segments within the voltage sensing domain (VSD),  as well as induced charge 
at the dielectric boundary of the membrane, the potential function
 for the S4 sensor is a nonlinear function of $Z$ for each potential $V$ \cite{llg,va2}.

By expressing a cloned \textit{Shaker} IR $K^+$ channel in \textit{Xenopus}
oocytes, contamination with endogenous currents is minimized, and permits 
a more accurate determination of the ionic and gating currents across the
membrane \cite{zha}. Assuming that channel opening occurs when the voltage
sensor in each of the four subunits is activated through two transitions
between three states, followed by a cooperative transition, the model
accounts for steady state and kinetic activation and deactivation for both
ionic and gating currents, the measurement of gating charge of 13e per
channel during activation, and the rising phase of the gating current
observed during a large depolarization.  

 Each of the four subunits of a voltage-dependent $K^+$ channel has a voltage-sensing
 domain with transmembrane segments S1 to S4, and  segments S5 to S6 that form 
the  ion-conducting pore domain. The  S4 segment  moves transversely through a gating pore
 in  response to a change in  the potential difference across the membrane,
 and is stabilized by the interaction between its positively charged residues R1 to
R4 and negatively charged  amino acids  on S2 and S3 segments \cite{tlsp,srp}, and 
in the activated state, lipid phosphate groups on the membrane surface \cite{sjm,jr,kjy}. Metal-ion constraints on the activation process and molecular dynamics simulations have 
identified at least three transitions between the resting state and activated state
 of the S4 sensor within a voltage sensing domain of a $K^+$ ion channel 
\cite{hrb,dtk,jjb}. The attached residues of the S4 segment  sequentially 
translocate across a focused membrane field where the hydrophobic residue 
$F^{290}$ on the S2 segment of the $K^+$ channel contributes to the energy barrier for
 transitions \cite{tlldm}. By measuring the effect of mutants on the gating current of the $K^+$ channel, it has been shown that $F^{290}$ controls the  transfer of the arginine 
R4 across the membrane field during the final gating transition, and has a significant effect on deactivation kinetics but only a small effect on activation time  \cite{lb}. 
 When the hydrophobic residue I287 on the S2 segment or V363 adjacent to the residue R1 on the S4 segment are replaced by the hydrophilic amino acid Thr, the rate of activation of the $K^+$ channel is increased by a factor of two, and comparable to the activation rate in a $Na^+$  ion channel \cite{lcfb}. Therefore, the first forward transition during activation and the first backward transition during deactivation  are rate-limiting, and dependent  on the hydrophobicity of residues on adjacent segments in the VSD.

Assuming that the energy of the voltage sensor is dependent on the Coulomb
force between positive S4 residues and negative residues on neighboring
segments, the electric field within the membrane and the dielectric boundary
force on charged residues, the energy of each potential well and, therefore,
the equilibrium distribution of gating charge for each voltage may be
calculated for both wild-type and mutant {\em Shaker}  $K^+$ channels, and
good agreement with experimental data is obtained by an appropriate choice
of the positions of the negative residues and of the other parameter values
\cite{llg}. The time-dependence of the survival probabilities of the closed states
 of the voltage sensor during a voltage clamp  may be 
calculated from a solution of Smoluchowski's equation
 with a potential function that may be derived by application of 
Poisson's equation to charged residues within a dielectric slab between solvent regions,
 and is in accord  with a three-stage empirical model of the ionic and gating currents for a 
$K^+$ channel \cite{va3,ss}. In particular,  the voltage dependence
 of the forward and backward rate functions for each stage may be derived from the
asymmetry of the energy barriers determined by the  interaction between  S4 residues 
in close proximity to residues  on S2 and S3 segments.

In this paper, assuming that the $K^+$ channel conductance is modulated by a single 
S4 sensor which activates in two sequential stages, it is shown that if the 
 rate functions satisfy $\delta \ll \beta$, and either  $\alpha \ll \gamma $ or
 $\alpha \gg \gamma$, the master equation that describes the dynamics of a voltage 
clamp may be reduced to a rate equation with derived forward and backward rate functions
that approximate the functions  $\alpha_{n}$ and  $\beta_{n}$
 for  a squid axon $K^+$ channel \cite{hh}.  A three-stage model of activation of a
 $K^+$ channel may be determined that has the  same derived rate functions 
and survival probability $n(t)$ during a voltage clamp  as a 
two-stage model. If $\delta \ll \beta$, and either  $\alpha \ll \gamma $ or
 $\alpha \gg \gamma$, the activation of a $K^+$ ion channel during an 
action potential may be described by a single variable  $n$ with transitions 
between two states. However, if the rate functions satisfy  $\alpha \ll \gamma $ and 
$\beta \ll \delta$, both low and high frequencies contribute to the voltage 
clamp, and for an appropriate choice of  parameters, the ionic current
 equation and the master equation for $K^+$ channel 
activation exhibit a mixed-mode or bursting oscillation. 

\newpage 

\textbf{THE VOLTAGE CLAMP OF A MULTI-STAGE $K^+$ CHANNEL SENSOR}

It is assumed that a  $K^+$ channel voltage sensor is a Brownian
particle  that translocates across barriers between potential
wells by absorbing  energy from the thermal environment \cite{llg}.  The 
solution of the Smoluchowski equation for a square-well approximation to
the potential function of  the S4 sensor satisfies a master equation \cite{va2},
 and if activation occurs in two stages between three states (see Fig.
1)  \cite{zha},  the occupation probabilities of the closed states $n_{1}$, $n_{2}$
 and the open state $n$ are determined by 
\begin{eqnarray}
\frac{dn_{1}(t)}{dt} & = & -\alpha n_{1}(t)+\beta n_{2}(t), \label{nn1} \\
\frac{dn_{2}(t)}{dt} & = & \alpha n_{1}(t)+\delta n(t)-(\beta +\gamma )n_{2}(t), \label{nn2} \\
\frac{dn(t)}{dt}     & = & \gamma n_{2}(t)-\delta n(t), \label{nn}
\end{eqnarray}
where the rate functions
\begin{equation}
\alpha (V)=\alpha _{0}\exp [q_{\alpha }(V-V_{0})/kT],\ \beta (V)=\beta
_{0}\exp [-q_{\beta }(V-V_{0})/kT],
\end{equation}
\begin{equation}
\gamma (V)=\gamma _{0}\exp [q_{\gamma }(V-V_{0})/kT],\ \delta (V)=\delta
_{0}\exp [-q_{\delta }(V-V_{0})/kT],
\end{equation}
$\alpha _{0},\beta _{0},\gamma _{0},\delta _{0}$ are constants (ms$^{-1}$), $
q_{i}$ is the equivalent charge for each forward or backward transition, $k$
is Boltzmann's constant, $kT/e\approx 25$ mV, $e$ is the electronic charge,
$V_{0}$ is a constant and $V$ is the membrane voltage (in mV).

For a large hyperpolarized holding potential, we may assume that $n_{1}(0)=1$,
and $n(0)=n_{2}(0)=0$, and if the $K^+$ channel voltage sensor is depolarized to a
clamp potential $V$ (activation), the solution of Eqs. (\ref{nn1}) to (\ref{nn}) 
for the open state is
(see Appendix 1)
\begin{equation}
n_{A}(t)=\frac{\alpha \gamma }{\omega _{1}\omega _{2}}+\frac{\alpha \gamma }{%
\omega _{1}(\omega _{1}-\omega _{2})}\exp (-\omega _{1}t)-\frac{\alpha
\gamma }{\omega _{2}(\omega _{1}-\omega _{2})}\exp (-\omega _{2}t),
\label{act}
\end{equation}
where $\omega _{1}$ and $\omega _{2}$ ($> \omega _{1}$) are solutions of
the characteristic equation 
\begin{equation}
\omega ^{2}-\omega (\alpha +\beta +\gamma +\delta )+\alpha \gamma +\delta
(\alpha +\beta )=0.  \label{freq}
\end{equation}
However, if the S4 sensor is initially in the open state 
($n(0)=1$, and $n_{1}(0)=n_{2}(0)=0$), and if the $K^+$ channel sensor is repolarized to a  clamp
potential $V$ (deactivation), the solution is (see Appendix 1)
\begin{equation}
n_{D}(t) = \frac{\alpha \gamma }{\omega _{1}\omega _{2}}+\frac{\delta
(\omega _{2}-\gamma -\delta )}{\omega _{1}(\omega _{2}-\omega _{1})}\exp
(-\omega _{1}t)  
 +\frac{\delta (\omega _{1}-\gamma -\delta )}{\omega _{2}(\omega
_{1}-\omega _{2})}\exp (-\omega _{2}t).  \label{deact}
\end{equation}

Assuming that the rate functions of the first step are larger than the
second step ( $\beta \gg \delta $ and $\alpha \gg \gamma $), it may be shown
from Eq. (\ref{freq}) that $\omega _{1}\approx \gamma \alpha /(\alpha
+\beta )+\delta $ \ and $\omega _{2}\approx \alpha +\beta \gg \omega _{1}$,
and Eqs. (\ref{act}) and (\ref{deact}) become
\begin{eqnarray}
n_{A}(t) & \approx & \frac{\alpha \gamma }{\alpha \gamma +\delta (\alpha +\beta )}
[ 1-\exp (-\omega _{1}t)] ,  \label{act2} \\
n_{D}(t)  & \approx & \frac{\alpha \gamma +\delta (\alpha +\beta )\exp (-\omega
_{1}t)}{\alpha \gamma +\delta (\alpha +\beta )}.  \label{deact2}
\end{eqnarray}
Eqs. (\ref{act2}) and (\ref{deact2}) are solutions of the rate equation \cite
{hh}
\begin{equation}
\frac{dn}{dt}= \alpha _{n,2}-(\alpha _{n,2}+\beta_{n,2})n,  \label{ratek}
\end{equation}
where
\begin{equation}
\alpha_{n,2}(V)=\frac{\omega _{1}\alpha \gamma }{\alpha \gamma +\delta
(\alpha +\beta )},  \label{alf}
\end{equation}
\begin{equation}
\beta_{n,2}(V)=\frac{\omega _{1}\delta (\alpha +\beta )}{\alpha \gamma
+\delta (\alpha +\beta )},  \label{bet}
\end{equation}
and, therefore, $\omega _{1}=\alpha _{n,2}+\beta _{n,2}$,  and  
$\beta_{n,2}/\alpha_{n,2}=(1+\beta /\alpha )\delta /\gamma $. Substituting
 $\omega_{1}\approx \gamma \alpha /(\alpha +\beta )+\delta $, we may write
\begin{eqnarray}
\alpha _{n,2}(V) & \approx & \frac{\gamma }{1+\beta /\alpha },  \label{alf2} \\
\beta _{n,2}(V)  & \approx & \delta ,  \label{bet2}
\end{eqnarray}
which provide a good fit to the empirical rate functions $\alpha_{n}$ and
 $\beta_{n}$ for the squid axon $K^+$ channel \cite{hh}  (see Fig. 2). The subscript
'n,2' for $\alpha$ and $\beta$ in Eqs. (\ref{alf2})  and (\ref{bet2})  denotes a two-stage expression for the $K^+$ ion channel rate functions.  
The rate functions are chosen so that $\gamma \approx \alpha_{n}$ for 
large depolarized potentials, the ratio $\beta /\alpha \approx \gamma/\alpha_{n}-1$, and
 $\delta = \beta_{n}$. There is also good agreement
between the survival probability of the state  ${\normalsize n(t)}$
during a voltage clamp calculated from a rate equation with
$\alpha _{n}$\ and $\beta _{n}$,  and from the master
equation solutions \ Eqs. (\ref{act}) \ and (\ref{deact}) (see Figs. 3(a)
and 3(b)), and therefore, an activation process  for the voltage sensor with at least two stages provides a physical basis for the mathematical form and asymmetry of
 $\alpha _{n}$\ and $\beta _{n}$, without assuming a constant electric 
field within the membrane.  The function $\alpha _{n,2}(V)$  has the
almost linear variation of $\gamma (V)$  for large depolarizing clamp
potentials (as $q_{\gamma} \approx 0.3 e$), and has the exponential variation of 
$\beta (V)$, as well as $\alpha (V)$ and $\gamma (V)$, for depolarizations near the
 resting state, whereas $\beta_{n,2}(V)$ has the exponential voltage dependence of $\delta (V)$.

However, if we assume that $\beta \gg \delta $ and $\gamma \gg \alpha $, a
limiting case of the description of ionic and gating currents within the 
\textit{Shaker} $K^+$ ion channel \cite{zha}, it may be shown that $\omega
_{1}\approx \alpha \gamma /(\gamma +\beta )+\delta (\alpha +\beta )/(\gamma
+\beta )$\ and $\omega _{2}\approx \gamma +\beta \gg \omega _{1}$, and 
Eqs. (\ref{act}) and (\ref{deact})  become
\begin{eqnarray}
n_{A}(t) & \approx & \frac{\alpha \gamma }{\alpha \gamma +\delta (\alpha +\beta )}
[ 1-\exp (-\omega _{1}t)] ,  \label{act3} \\
n_{D}(t)  & \approx & \frac{\alpha \gamma }{\alpha \gamma +\delta (\alpha +\beta )}%
+\frac{\delta (\gamma -\alpha )\exp (-\omega _{2}t)}{(\gamma +\beta )^{2}}+%
\frac{\delta (\beta -\delta )\exp (-\omega _{1}t)}{(\gamma +\beta )(\alpha
+\delta )},
\end{eqnarray}
and for a sufficiently large hyperpolarizion, $\beta -\delta >\gamma
-\alpha$, and we may write
\begin{equation}
n_{D}(t)\approx \frac{\alpha \gamma +\delta (\alpha +\beta )\exp (-\omega
_{1}t)}{\alpha \gamma +\delta (\alpha +\beta )}.  \label{deact3}
\end{equation}
Eqs. (\ref{act3}) and (\ref{deact3}) are also solutions of Eq. (\ref{ratek}),
  and substituting $\omega _{1}=\alpha \gamma /(\gamma
+\beta )+\delta (\alpha +\beta )/(\gamma +\beta )$  into  Eqs.
(\ref{alf}) \ and (\ref{bet}),
\begin{eqnarray}
\alpha_{n,2}(V) & \approx & \frac{\alpha }{1+\beta /\gamma }  \label{alf3} \\
               & = & \frac{\alpha _{0}\exp [q_{\alpha }V/kT]}{1+
           (\beta _{0}/\gamma_{0})\exp [-(q_{\beta }+q_{\gamma })V/kT]}, \nonumber \\ 
\beta_{n,2}(V) & \approx  & \frac{\delta (\alpha +\beta )}{(\gamma +\beta )}  \label{bet3} \\
               & = &\frac{\delta_{0}\exp [-q_{\delta }V/kT](1+(\alpha _{0}/\beta_{0})
\exp [(q_{\alpha }+q_{\beta })V/kT])}{1+(\gamma _{0}/\beta_{0})
\exp [(q_{\gamma }+q_{\beta })V/kT]},  \nonumber
\end{eqnarray}
if $V_0 = 0$. For a large depolarizing potential $V$,  $\alpha _{n,2}(V)\approx \alpha $
and $\beta _{n,2}(V)\approx \delta \alpha /\gamma $, whereas for a hyperpolarizing potential,
  $\alpha _{n,2}(V)\approx \gamma \alpha /\beta $ and $\beta _{n,2}(V)\approx
\delta$, and have a similar form to the empirical rate functions for a
delayed rectifier $K^+$ ion channel in a cardiac Purkinje fiber  \cite{dn}, 
which may be expressed as
\begin{equation}
\alpha _{x}=\frac{D\exp [C(V-A)]}{1+\exp [-B(V-A))]},
\end{equation}
\begin{equation}
\beta _{x}=\frac{H\exp [-G(V-E))]}{1+\exp [F(V-E))]},
\end{equation}
where A to H are constants.

For a \textit{Shaker} $K^+$ ion channel \cite{zha}, the derived rate
functions $\alpha _{n,2}$ and $\beta _{n,2}$ \ may be approximated by functions $
\alpha_{H}$ and  $\beta_{H}$ which have the same mathematical form as the
 rate functions $\alpha _{n}$\ and $\beta _{n}$
(see Fig. 4). Although  there is good agreement
for small depolarizations between the survival probability $ n(t)$
 during a voltage clamp calculated from a
rate equation with $\alpha _{n,2}$\ and $\beta _{n,2}$, and
from Eqs. (\ref{act})  and (\ref{deact}) (see Figs. 5(a) and 5(b)), for
larger depolarizations the higher frequency component of the solution makes
a contribution, and therefore deviates from the two-state model. If $\beta
(V)$ and  $\gamma (V)$ are increased by a factor of 3, the rate equation
provides a better fit to two-stage activation for larger depolarizations
(see Fig. 5(c)). However,  the rate equation can
not account for the rising phase of the K channel gating current that has
been observed in both \textit{Shaker}  and squid axon K channels \cite{zha},
and requires an activation model of the ion channel for which 
$\gamma > \alpha $ (see Appendix 2).

If the rate functions satisfy $\beta \ll \delta $ and $\alpha \ll \gamma $,
from Eqs.\ (\ref{act}) and (\ref{freq}), $\omega_{1}\approx \alpha
+\beta \delta /(\gamma +\delta ) \ll \omega _{2}\approx \gamma +\delta $,
and
\begin{equation}
n_{A}(t)\approx \frac{\alpha \gamma }{\omega _{1}\omega _{2}}+\frac{\alpha
\gamma }{\omega _{2}}\left( \frac{-\exp (-\omega _{1}t)}{\omega _{1}}+ \frac{
\exp (-\omega _{2}t)}{\omega _{2}}\right) .
\end{equation}
The high frequency term is small for all potentials, and hence
\begin{equation}
n_{A}(t)\approx \frac{\alpha \gamma }{\alpha \gamma +\delta (\alpha +\beta )}
[1-\exp (-\omega _{1}t)].
\end{equation}
For deactivation, \ the low frequency term in Eq.\ \ (\ref{deact}) \ is
small, and
\begin{equation}
n_{D}(t)\approx \frac{\alpha \gamma +\delta (\alpha +\beta ) + 
\delta \omega _{1}(\exp (-\omega_{2}t) - 1)}{\alpha \gamma +\delta (\alpha +\beta )}.
\end{equation}
Therefore, the ionic currents during a voltage clamp cannot be described by a rate equation, and similarly, if $\beta \ll \delta $ and $\alpha \gg \gamma $, the ionic
currents have contributions from both low and high frequency terms, and
therefore require a master equation description.

Based on the measurement of macroscopic ionic and gating currents and single channel currents across a $K^+$ ion channel membrane over a broad voltage range, 
it has been proposed that the voltage sensor in each of the four subunits is activated through three transitions before ion channel opening \cite{ss},
 and has received support from molecular dynamics simulations
 of the voltage sensor domain within $K^+$ ion channels \cite{hrb,dtk,jjb}.
If the S4 voltage sensor is activated in three stages between four states 
(see Fig. 6), the solution of the Smoluchowski equation for the activation  landscape may be expressed as a master equation
\begin{eqnarray}
\frac{dn_{1}(t)}{dt} & = & -\alpha_1 n_{1}(t)+\beta_1 n_{2}(t),  \label{na1} \\
\frac{dn_{2}(t)}{dt} & = & \alpha_1 n_{1}(t)+\beta_2 n_{3}(t)-(\beta_1 +\alpha_2)n_{2}(t),
\label{na2} \\
\frac{dn_{3}(t)}{dt} & = & \alpha_2 n_{2}(t)+\beta_3 n(t) - (\beta_2 +\alpha_3 )n_{3}(t),
\label{na3} \\
\frac{dn(t)}{dt}     & = & \alpha_3 n_{3}(t)-\beta_3 n(t),  \label{na}
\end{eqnarray}
where the rate functions
\[
\alpha_i(V)= \alpha _{i0}\exp [q_{\alpha_i}V/kT],
 \beta_i(V)= \beta_{i0}\exp [-q_{\beta_i}V/kT],
\]
for $i = 1$ to $3$, $\alpha_{i0}$ and $\beta_{i0}$ are constants (ms$^{-1}$),  and $q_{\alpha_i}$  and $q_{\beta_i}$ are  the equivalent charges for the $i$th forward or backward transition for each stage. 

Assuming that the rate functions for the last step of activation of the S4 sensor across the membrane are an order of magnitude smaller than previous steps 
 ( $\beta_1 \gg \beta_2 \gg \beta_3 $ and 
$\alpha_1 \gg \alpha_2 \gg \alpha_3 $), the master equation may be reduced to
 (see Appendix 1)
\begin{equation}
\frac{dn}{dt}= \alpha_{n,3}-(\alpha_{n,3}+\beta_{n,3})n,  \label{rate3st}
\end{equation}
where the three-stage expressions for the $K^+$ channel rate functions (denoted by the subscript 'n,3')  are
\begin{eqnarray}
\alpha _{n,3}(V) & \approx & \frac{\alpha_3}{1 + (\beta_2/\alpha_2)(1 + \beta_1/\alpha_1)},
\label{alfn3} \\
\beta _{n,3}(V)  & \approx & \beta_3,  \label{betn3}
\end{eqnarray} 
which also provide a good fit to the rate functions  $\alpha_{n}$ and  $\beta_{n}$ 
for the squid axon $K^+$ channel, and reduce to the expressions in Eqs. (\ref{alf2}) and  (\ref{bet2}) when $\beta_1 = 0$.
However, if it is assumed that the rate functions satisfy $\beta_3 \ll \beta_1, \beta_2$ and $\alpha_1 \ll \alpha_2, \alpha_3$, the master equation may be approximated by  Eq.  (\ref{rate3st}) where
\begin{equation}
\alpha_{n,3}(V) \approx
\frac{\alpha_1 \alpha_3}{\alpha_3 + (\beta_2/\alpha_2)(\alpha_3 \beta_1/\beta_2 + \beta_1)},  \label{alf3st}
\end{equation}
\begin{equation}
\beta_{n,3}(V) \approx \frac{\beta_3 (\alpha_1 + \beta_2 (\alpha_1 + \beta_1)/\alpha_2)}
{\alpha_3 +  (\beta_2/\alpha_2)(\alpha_3 \beta_1/\beta_2 + \beta_1)
 }.  \label{bet3st}
\end{equation}
For a large depolarizing potential $V$,  $\alpha_{n,3}(V) \approx \alpha_1 $
and $\beta_{n,3}(V) \approx \beta_3 \alpha_1/\alpha_3$,
 whereas for a hyperpolarizing potential,  $\alpha_{n,3}(V) \approx  \alpha_1 (\alpha_2 \alpha_3)/(\beta_1 \beta_2)$  and $\beta_{n,3} (V)$  $\approx \beta_3 $.

If a two and three stage model of activation of a voltage sensor each reduce to the same rate equation, $\alpha_{n,3}(V) = \alpha_{n,2}(V)$ and $\beta_{n,3}(V) = \beta_{n,2}(V)$, and  assuming $\alpha_1 \approx \alpha_3 \beta_1/\beta_2$, the rate functions   $\alpha = \alpha_1$, $\delta = \beta_3$, $\gamma = \alpha_3$,     
$\beta \approx \beta_2(\alpha_1 + \beta_1)/\alpha_2 \approx   \beta_2(\alpha_3 \beta_1/\beta_2 + \beta_1)/\alpha_2$, and hence the effective charge $q_{\beta} \approx q_{\beta_1} + 
q_{\beta_2} - q_{\alpha_2}$ has contributions from several transitions. That is, a 
three-stage model of activation of a $K^+$ ion channel sensor may be determined 
that has the  same derived rate functions and survival probability $n(t)$ during a voltage clamp  as a 
two-stage model (see Figs. 7 and 8). The three-stage rate functions 
have a similar voltage dependence to those obtained from an electrostatic model
 of S4 activation \cite{va3}, where  approximately $1e$ is transferred across the membrane for each of the three activation steps per subunit, and provides a good fit to an empirical three-stage model of $K^+$ channel activation \cite{ss}.

\newpage 
\textbf{TWO-STAGE $K^+$ ION CHANNEL SENSOR ACTIVATION  AND THE ACTION
POTENTIAL}

In this section, we consider the effect of two-stage activation of a K
channel on potential oscillations across a membrane. The \textit{Shaker} $K^+$
channel permits the conduction of ions when the activation of each of
several voltage sensors through a sequence of closed states is followed by a
cooperative transition to the open state \cite{zha}. However, in order to
simplify the analysis, it is  assumed that (1) the $K^+$ conductance  is
modulated by a single S4 sensor that activates in two stages (2) each
ion channel is persistent, that is, does not inactivate and (3) the activation 
of the $Na^+$ channel sensor is instantaneous. A more general analysis 
may take account of the effect of  relaxing each of these assumptions 
on the membrane potential and the bifurcations of the system of equations.

The potential response of a membrane with non-inactivating $Na^+$, $K^+$ and leakage ion channels may
be described by the current equation
\begin{equation}
C\frac{dV}{dt}=I-\overline{g}_{Na}m_{\infty }(V)(V-V_{Na})-\overline{g}
_{K}n(V-V_{K})-\overline{g}_{L}(V-V_{L}),  \label{cur1}
\end{equation}
and the master equation for two-stage activation of the $K^+$ ion channel sensor
\begin{equation}
\frac{dn}{dt}=\phi _{n}(V) \left( \frac{1-n_{1}}{1+\rho _{n}\exp [-(q_{\gamma
}+q_{\delta })(V-V_{0})/kT]}-n \right),  \label{n}
\end{equation}
\begin{equation}
\frac{dn_{1}}{dt}=\phi _{1}(V) \left( \frac{1-n}{1+\rho _{1}\exp [(q_{\alpha
}+q_{\beta })(V-V_{0})/kT]}-n_{1} \right),  \label{n1}
\end{equation}%
where Eq. (\ref{nn2}) has been eliminated by application of $n_{2}=1-n_{1}-n$, 
$\overline{g}_{i}$ is the maximal ion conductance and $V_{i}$ is the equilibrium
potential for each ion ($Na^+$ , $K^+$ and leakage), $\rho _{1}=\alpha _{0}/\beta _{0}
$, $\rho _{n}=\delta _{0}/\gamma _{0}$, $\phi _{1}(V)=\alpha (V)+\beta (V)$, 
$\phi _{n}(V)=\gamma (V)+\delta (V),$ C=1 $\mu $F/cm$^{2}$. The rate functions
 for $Na^+$ channel activation in the squid axon are
\begin{eqnarray}
\alpha_{m} & = & \frac{0.1(V+V_{m}-25)}{1-\exp [-0.1(V+V_{m}-25)]},\nonumber \\
\beta_{m} & = & 4\exp[-(V+V_{m})/18], \label{alfm}
\end{eqnarray} 
where $V_{m}$ is a parameter, and  the steady state $Na^+$  ion channel open
 probability is\ $m_{\infty }(V)=\alpha _{m}/(\alpha _{m}+\beta _{m})$,
 a good approximation to the  activation variable\ $m$ when the $Na^+$  ion channel
activation time is small.

Based on the effect on $Na^+$  ion channel currents of double-cysteine mutants of
S4 gating charges and the negative charge E43 on the S1 segment, structural models
of resting and activated states of the VSD of the bacterial $Na^+$  channel NaChBac,
which resembles a vertebrate $Na_V$ ion channel domain,
demonstrate that at least two transitions occur during activation of each voltage sensor \cite{dystc}. This conclusion is consistent with a model of the activation
of a  Na channel based on the measurement of a rising phase of the gating current in a 
squid axon, and the chemical structure of a Na channel  \cite{key}.
 Therefore, adopting the two-stage expressions in Eqs (\ref{alf2}) and (\ref{bet2}) for the $Na^+$  ion channel rate functions (denoted by the subscript 'm,2'), we may define
$\alpha_{m,2}=2.6\exp [0.28(V+V_{m})/25]/(1+ 8.4 \exp [-z_m(V+V_{m})/25])$, where $z_m$
is a parameter, such that $\alpha_{m,2} \approx \alpha_{m}$ when $z_m = 1.3$, 
$\beta _{m,2} = \beta_{m}$, and
$m_{\infty }(V) \approx \alpha _{m,2}/(\alpha _{m,2}+\beta _{m,2})$. 
The empirical $Na^+$  and $K^+$ ion channel forward rate functions for the squid axon ($\alpha_{m}$ and $\alpha_{n}$) have  a similar mathematical form because their respective voltage sensors require more than one step to activate. 

The stationary points of the system of Eqs. (\ref{cur1}) - (\ref{n1}) 
are given by the intersection of the $V$, $n$,  and $n_{1}$ nullclines,
and their stability may be determined from the characteristic equation (see
Appendix 3). If the two-stage activation of the $K^+$ ion channel satisfies 
$\beta \gg \delta $, and $\alpha \gg \gamma $, we may substitute 
 $n_{1} = (1-n)\beta/(\alpha +\beta)$ into Eq. (\ref{n}) to obtain a rate equation for $n$,
 and hence the  solutions of Eqs. (\ref{ratek}), 
 (\ref{alf2}), (\ref{bet2}) and (\ref{cur1}) provide a good fit to the solutions
 of Eqs. (\ref{cur1}) - (\ref{n1})  - see Figs. 9 and 10 for the squid axon $K^+$ ion channel,
 where the bifurcation analysis is generated by XPPAUTO  \cite{erm}. However, if
$\beta \gg \delta $, and $\alpha \ll \gamma$, we may substitute  
$n_{2} = (\alpha n_{1} +\delta n)/(\beta + \gamma)$ into Eqs. (\ref{nn1}) and 
(\ref{nn}) to obtain Eq. (\ref{ratek}) where
\begin{eqnarray}
\alpha_{n,2}(V) & \approx & \frac{\alpha \gamma}{\alpha  + \gamma + \beta},  \label{alf4} \\    
\beta_{n,2}(V) & \approx  & \frac{\delta (\alpha +\beta )}{\alpha + \gamma +\beta },  \label{bet4}    
\end{eqnarray}
which reduce to Eqs. (\ref{alf3}) and (\ref{bet3}) when $\gamma \gg \alpha$, and therefore, the  solutions of Eqs. (\ref{ratek}), (\ref{cur1}), (\ref{alf4}) and
 (\ref{bet4})  provide a good fit to the solutions of Eqs. (\ref{cur1}) - (\ref{n1}) - see Figs. 11 and 12 for the \textit{Shaker} $K^+$ ion channel.
 If $q_{\gamma}, q_{\delta} \ll e$, the range of currents for which there is a coexistence of a stable state and a stable limit cycle (bistability)  is small (see Fig. 10). However, for larger values of $q_{\delta}$, similar to measured values for the \textit{Shaker} $K^+$ ion channel, the range of bistability increases for the bifurcation point HB1 (see Fig. 12), and for  values of $q_{\alpha} \approx q_{\gamma}$ greater than the normal range for $K^+$ ion channels ($ > 0.4 e$),
there may exist a region of bistability for each bifurcation point.

However, if $\beta \ll \delta $, and $\alpha \ll \gamma $, the variation in the
probability $n_{1}$ that the sensor occupies the first closed state is an
order of magnitude slower than for the variable $n$, and hence $n_{1}$ may
be treated as a parameter that modifies the stability of the stationary state in 
the (V, n) subsystem (see Fig. 13)  \cite{rin}. During the subthreshold
 oscillation,  from Eq. (\ref{n1}) $n_{1}$ increases until the stationary state in
 the subsystem becomes unstable, and the  trajectory spirals towards an action 
potential burst (see Figs. 13 and 14). However, during the large  amplitude oscillation,  $n_{1}$ progressively decreases until the stationary state is stable 
and the system returns to equilibrium, followed by a loss of stability as 
$n_{1}$ again increases, and therefore,  Eqs. (\ref{cur1}) -  (\ref{n1})
may exhibit a mixed-mode or bursting oscillation.

\newpage 
\textbf{CONCLUSION}

A two-stage model of the activation of a voltage sensor in each of the four subunits 
 of a \textit{Shaker} K+ ion channel, followed by a cooperative transition,  can describe the
kinetic activation and deactivation of both ionic and gating currents \cite{zha},
 but does not account for the success of the two-state 
Hodgkin-Huxley  model \cite{hh}. In this paper, we show that if the rate
functions satisfy $\delta \ll \beta $ and either $\alpha \gg \gamma $ or  
$\gamma \gg \alpha $, the solution of a two-stage model of the activation of an
ion channel sensor during a voltage clamp may be approximated by the solution of a
rate equation where the backward transition rate is an exponential function
of $V$, and the forward rate may be expressed as an exponential-quasilinear
function of $V$, and have a similar form to the empirical rate functions $
\alpha_{n}$\ and $\beta_{n}$ of the squid axon $K^+$ ion channel, and the
functions $\alpha_{x}$ and $\beta_{x}$ of a delayed rectifier $K^+$ ion channel
in a cardiac Purkinje fiber {\normalsize \cite{hh, dn}}. 

The derived rate functions account for the asymmetry of  $\alpha_{n}$ and 
$\beta_{n}$ in terms of the effective charge for the transitions of the
activation process, which may be calculated from  the voltage-dependence 
of the barrier heights for each stage of activation of the S4 sensor \cite{va3}.
If the opening of a $K^+$ ion channel occurs upon the activation of each of four 
independent two-stage voltage sensors,  the $K^+$ conductance may be expressed as 
$\overline{g}_{K}n^4$, where $n$ is the solution to the corresponding
 rate equation with derived rate functions $\alpha_{n,2}$ and $\beta_{n,2}$.
However, during the voltage clamp of a two-stage opening of a  $K^+$ ion channel,
 the gating current has a contribution from each transition and  exhibits a rising phase
 for a large depolarization  when  $\gamma >\alpha $, and therefore, is in accord
 with experimental data from  \textit {Shaker} \cite{zha} and squid axon \cite{wb}
 $K^+$ ion channels.

 If the S4 sensor is activated through three transitions where the first forward transition  and the first backward transition  are rate limiting
($\alpha_1 \ll \alpha_2, \alpha_3$ and $\beta_3 \ll \beta_1, \beta_2$), 
 the master equation during a voltage clamp also reduces  to a rate equation. 
By expressing a two-stage model of voltage sensor activation as a  three-stage model with the same  derived rate functions and survival probability for the activated state, it may be shown that an empirical model of $K^+$ ion channel activation \cite{zha}
is consistent with models based on the structure of a voltage sensing domain in which 
approximately $1e$ is transferred across the membrane for each of the three activation 
steps per subunit. The  rate function inequalities are supported by  recent experiments that have shown that the activation time is reduced when the residue V363 adjacent to R1 on the S4 segment and the residue I287 on the S2 segment are replaced by the hydrophilic amino acid Thr \cite{lcfb}, and that the deactivation time for the first backward transition is dependent on the hydrophobicity of the residue $F^{290}$ on the S2 segment of a  $K^+$ ion channel subunit  \cite{tlldm,lb}. 

 Two-stage activation of the $K^+$ ion channel during an action potential
 may be described by a single 
variable $n$ when $\delta \ll \beta $ and either $\alpha \gg \gamma $ or  
$\gamma \gg \alpha $, and  therefore, is consistent with experimental data for the squid axon $K^+$ ion channel \cite{hh}. However, if  $\beta \ll \delta $ and  $\alpha \ll \gamma$,
 the system may exhibit a mixed-mode or bursting  oscillation, in which the alternation of  repeated spikes with a subthreshold oscillation may be attributed to the  modulation of the $K^+$ conductance by the variation in the occupation probability of the inner closed state $n_{1}$.

The interaction between inactivating $K^+$ ion channels and $Na^+$  ion channels may also
generate bursting in  fast spiking cortical neurons but, in this case,
deinactivation increases the $K^+$ conductance until the burst can not be
sustained, and inactivation reduces the $K^+$ conductance during the quiescent
phase enabling another burst of spikes \cite{wa,rr}. Bursting may also occur
when a slow M-type $K^+$ current modulates the repetitive action potential
generated by $Na^+$  ion channels and delayed rectifier $K^+$ ion channels, and has been
observed in hippocampal CA1 pyramidal neurons \cite{yy}. Each of these
examples of bursting incorporates an additional variable or ionic current
that modifies the stability of the stationary state, but the HH model can
exhibit a mixed-mode oscillation when the physiological time constant for
either the $K^+$ channel activation variable $n$ or the $Na^+$ channel inactivation variable $h$ is
increased by an order of magnitude \cite{dnk,rw}. Similarly, the equations
that describe the interaction between a persistent $Na^+$  ion channel and the
two-stage opening of a $K^+$ ion channel have bursting solutions that are suppressed
by the parameter values that characterize the $K^+$ ion channel rate functions.

\newpage 
\textbf{APPENDIX 1}

The master equation  for two-stage activation, Eqs. (\ref{nn1}) to (\ref{nn}),
 may be expressed as
\begin{eqnarray}
\frac{dn_{1}(t)}{dt} & = & -(\alpha + \beta) n_{1}(t)+\beta (1 - n(t)), \nonumber \\ 
\frac{dn(t)}{dt}     & = & \gamma (1 - n_{1}(t)) - (\gamma + \delta) n(t),  \label{e}
\end{eqnarray}
where Eq. (\ref{nn2}) has been eliminated by substitution of $n_2 = 1 - n_1 - n$.
The general solution of Eqs.  (\ref{e}) is
\begin{eqnarray*}
n_{1}(t) & = & n_{1s} + \sum_{i=1}^2 a_i(\omega_i - \gamma - \delta) \exp (-\omega_i t) \\
n(t)     & = & n_{s} + \gamma \sum_{i=1}^2 a_i \exp (-\omega_i t),
\end{eqnarray*}
where $n_{1s} = \beta \delta/(\alpha \gamma +\delta (\alpha +\beta ))$, $n_{s} = \alpha \gamma/(\alpha \gamma +\delta (\alpha +\beta ))$, $\omega _{1}$ and 
$\omega _{2}$ ($> \omega _{1}$) are solutions of
\[
\omega ^{2}-\omega (\alpha +\beta +\gamma +\delta )+\alpha \gamma +\delta
(\alpha +\beta )=0,
\]
and $a_1, a_2$ are constants determined by the initial conditions.
For a large hyperpolarized holding potential, we may assume that $n_{1}(0)=1$,
 and $n(0)=n_{2}(0)=0$, and if a $K^+$ channel voltage sensor is depolarized to a
clamp potential $V$,  by application of  the
initial condition, the solution of Eqs. (\ref{nn1}) to (\ref{nn}) for two-stage activation is
\[
n_{1A}(t)=\frac{\beta \delta }{\omega _{1}\omega _{2}}-\frac{\alpha (\gamma
+\delta -\omega _{1})}{\omega _{1}(\omega _{1}-\omega _{2})}\exp (-\omega
_{1}t)+\frac{\alpha (\gamma +\delta -\omega _{2})}{\omega _{2}(\omega
_{1}-\omega _{2})}\exp (-\omega _{2}t), \nonumber
\]
\[
n_{2A}(t)=\frac{\alpha \delta }{\omega _{1}\omega _{2}}-\frac{\alpha (\omega
_{1}-\delta )}{\omega _{1}(\omega _{1}-\omega _{2})}\exp (-\omega _{1}t)+%
\frac{\alpha (\omega _{2}-\delta )}{\omega _{2}(\omega _{1}-\omega _{2})}%
\exp (-\omega _{2}t),\nonumber
\]
\[
n_{A}(t)=\frac{\alpha \gamma }{\omega _{1}\omega _{2}}+\frac{\alpha \gamma }{%
\omega _{1}(\omega _{1}-\omega _{2})}\exp (-\omega _{1}t)-\frac{\alpha
\gamma }{\omega _{2}(\omega _{1}-\omega _{2})}\exp (-\omega _{2}t).
\]

However, if the S4 sensor is initially in the open state 
($n(0)=1$, and $n_{1}(0)=n_{2}(0)=0$), and if the $K^+$ sensor is repolarized to a  clamp
potential $V$, the solution is
\[
n_{1D}(t) = \frac{\beta \delta }{\omega _{1}\omega _{2}}-\frac{\beta \delta 
}{\omega _{1}(\omega _{2}-\omega _{1})}\exp (-\omega _{1}t) 
-\frac{\beta \delta }{\omega _{2}(\omega _{1}-\omega _{2})}\exp (-\omega
_{2}t),
\]
\[
n_{2D}(t) = \frac{\alpha \delta }{\omega _{1}\omega _{2}}+\frac{\delta
(\omega _{1}-\alpha )}{\omega _{1}(\omega _{2}-\omega _{1})}\exp (-\omega
_{1}t) 
+\frac{\delta (\omega _{2}-\alpha )}{\omega _{2}(\omega _{1}-\omega _{2})}%
\exp (-\omega _{2}t),
\]
\[
n_{D}(t) = \frac{\alpha \gamma }{\omega _{1}\omega _{2}}+\frac{\delta
(\omega _{2}-\gamma -\delta )}{\omega _{1}(\omega _{2}-\omega _{1})}\exp
(-\omega _{1}t)  
 +\frac{\delta (\omega _{1}-\gamma -\delta )}{\omega _{2}(\omega
_{1}-\omega _{2})}\exp (-\omega _{2}t). 
\]

The general solution of the three-stage master equation Eqs. (\ref{na1}) - (\ref{na})  is
\begin{eqnarray}
n_1(t) & = &  \frac{\beta_1 \beta_2 \beta_3 }{\omega_{1}\omega_{2}\omega_{3}}
+ \Sigma_{i = 1}^3  a_i g_1(\omega_i) \exp(-\omega_i  t), \nonumber \\ 
n_2(t) & = &  \frac{\alpha_1 \beta_2 \beta_3 }{\omega_{1}\omega_{2}\omega_{3}}
- \alpha_1 \Sigma_{i = 1}^3  a_i g_2(\omega_i) \exp(-\omega_i  t), \nonumber \\ 
n_3(t) & = &  \frac{\alpha_1 \alpha_2 \beta_3 }{\omega_{1}\omega_{2}\omega_{3}}
- \alpha_1 \alpha_2 \Sigma_{i = 1}^3  a_i (\beta_3 - \omega_i )\exp(-\omega_i  t), \nonumber \\ 
n(t) & = &   \frac{\alpha_1 \alpha_2 \alpha_3 }{\omega_{1}\omega_{2}\omega_{3}}
- \alpha_1 \alpha_2 \alpha_3 \Sigma_{i = 1}^3  a_i \exp(-\omega_i  t), \label{n34} 
\end{eqnarray}
where  $a_1,a_2$ and $a_3$ are constants, $g_1(\omega) = \omega^2 - (\alpha_2 + \alpha_3 + \beta_2 + \beta_3)\omega + \alpha_2(\alpha_3 + \beta_3)+ \beta_2 \beta_3$, $g_2(\omega) = \omega^2 - (\alpha_3 + \beta_2 + \beta_3)\omega +  \beta_2 \beta_3$, the frequencies
 $\omega_1 < \omega_2 < \omega_3$ are solutions of
\begin{equation}
\omega^3 - e_1 \omega^2 + e_2 \omega - e_3 = 0, 
\end{equation}
and
\begin{eqnarray*}
e_1 & = & \Sigma_{i=1}^3 (\alpha_i + \beta_i) =  \Sigma_{i=1}^3 \omega_i \\
e_2 & = & \alpha_1(\alpha_2 + \alpha_3 + \beta_2 + \beta_3) + \alpha_2(\alpha_3 + \beta_3) +  \beta_2\beta_3
 + \beta_1(\alpha_3 + \beta_2 + \beta_3) \\
    & = & \omega_1 \omega_2 + \omega_2 \omega_3 + \omega_1 \omega_3 \\
e_3 & = & \alpha_1 \alpha_2 \alpha_3  + \beta_3(\alpha_1 \alpha_2  + 
 \beta_2(\alpha_1 + \beta_1))  = \omega_1 \omega_2 \omega_3. \\
\end{eqnarray*}
If the $K^+$ channel voltage sensor is depolarized to a clamp potential $V$ from a
 hyperpolarized potential ($n_{1}(0)=1$ , and $n(0)=n_{2}(0)=n_{3}(0)=0$),
it may be shown that  $a_i = 1/(\omega_i \Pi_{j \neq i} (\omega_j -\omega_i))$.
If $\alpha_1 \gg \alpha_2 \gg \alpha_3 $ and $\beta_1 \gg \beta_2 \gg \beta_3$, 
the frequency $\omega_{1} \ll \omega_{2} \ll \omega_{3}$, 
\[
\omega_2 + \omega_3  \approx \Sigma_{i=1}^2 (\alpha_i + \beta_i),
\]
\[
\omega_2 \omega_3  \approx   \alpha_1 \alpha_2 + \alpha_1 \beta_2 +  \beta_1 \beta_2,
\]
\[
\omega_1 =   \alpha _{n,3} + \beta _{n,3}  =   \frac{e_3}{\omega_2 \omega_3}, 
\]
and $a_1 \gg a_2 \gg a_3$. That is, the  higher frequency terms 
 have small amplitude and $n(t)$ satisfies a 
rate equation with time constant $\omega_{1} = \alpha _{n,3}(V) + \beta _{n,3}(V)$
 where 
\begin{eqnarray*}
\alpha_{n,3}(V) & \approx & \frac{\alpha_3}{1 + (\beta_2/\alpha_2)(1 + \beta_1/\alpha_1)},
 \\
\beta_{n,3}(V)  & \approx & \beta_3.  
\end{eqnarray*}
However, if $\alpha_1 \ll \alpha_2, \alpha_3 $ and $\beta_3 \ll \beta_1, \beta_2$, 
\begin{eqnarray*}
\omega_2 + \omega_3  & \approx &  \Sigma_{i=1}^2 (\alpha_{i+1} + \beta_i),\\
\omega_2 \omega_3  & \approx &    \alpha_3  \alpha_2 + \alpha_3  \beta_1 + \beta_1 \beta_2,
\end{eqnarray*}
and the rate functions 
\[
\alpha_{n,3}(V) \approx
\frac{\alpha_1 \alpha_3}{\alpha_3 + (\beta_2/\alpha_2)(\alpha_3 \beta_1/\beta_2 + \beta_1)},  
\]
\[
\beta_{n,3}(V) \approx \frac{\beta_3 (\alpha_1 + \beta_2(\alpha_1 + \beta_1)/\alpha_2)}
{\alpha_3 +  (\beta_2/\alpha_2)(\alpha_3 \beta_1/\beta_2 + \beta_1)
 }.  
\]

\newpage 
\textbf{APPENDIX 2}

Although the $K^+$ ionic current is a function of the activation variable $n$
if the rate functions satisfy $\delta \ll \beta $ and either $\alpha \gg \gamma $ or  
$\gamma \gg \alpha $,
the gating current of the ion channel is comprised of a fast
component, considered to be a redistribution of the charge of the S4 sensor
within energy wells, and a slow component generated by transitions between
stationary states. The gating charge $\overline{Q}_{i}$ associated with the
transition to the i-th state from the first state is the change in charge
induced in the surrounding solvent \cite{roux}, and may be expressed as  $
\sum_{j=2}^{i}Q_{j}$ where $Q_{j}$ is the  gating charge for  the
transition from the  (j -- 1)th \ state to the j-th state.  The
observable gating charge $Q_{g}(t)$\ for each subunit of the $K^+$ ion channel is
dependent on the survival probability for each state, and if the voltage
 sensor activates in two steps,  $Q_{g,2}(t)=Q_{2}n_{2}(t)+(Q_{2}+Q_{3}) n(t),$
 and the gating current 
\begin{equation}
I_{g,2}(t)=Q_{2}\frac{dn_{2}}{dt}+(Q_{2}+Q_{3})\ \frac{dn}{dt}.
\label{gating}
\end{equation}

From  the solution of Eqs. (\ref{nn1}) to (\ref{nn}) in Appendix 1, we may write
\begin{eqnarray}
I_{g,2}(t) &=&\frac{\alpha (\gamma Q_{3}-(\alpha +\beta )Q_{2})}{
(\omega _{2}-\omega _{1})}[\exp (-\omega_{1}t) -  \exp(-\omega_{2}t)] +  \nonumber \\
&&\frac{\alpha Q_{2}}{(\omega _{2}-\omega _{1})}
[\omega_{2} \exp (-\omega_{1}t) - \omega_{1}\exp (-\omega_{2}t)], \label{rate ig}
\end{eqnarray}
and for $t \gg 1/\omega_2$, the high frequency terms are small when 
$\omega_{2} \gg \omega_{1}$,  and thus
\begin{equation}
I_{g,2}(t) \approx \frac{\alpha  \exp (-\omega _{1}t)}
{(\omega_{2}-\omega_{1})}\left[\gamma Q_{3} + (\omega_{2} - \alpha - \beta )Q_{2}  \right].
\label{gating2}
\end{equation}
If $\alpha \gg \gamma$ and $\beta \gg \delta$, $\omega_{1} \approx \alpha_{n,2} +\beta_{n,2}$, 
$\omega_{2} \approx \alpha +\beta$, and
\begin{equation}
I_{g,2}(t) \approx   \alpha_{n,2}  Q_{3} \exp (-\omega_{1}t),
\label{gating2a}
\end{equation}
where $\alpha_{n,2}$ is given by Eq. (\ref{alf2}).  However, if
 $\alpha \ll \gamma$ and $\beta \gg \delta$, $\omega_{1} \approx \alpha_{n,2} +\beta_{n,2}$, 
 $\omega_{2} \approx \gamma +\beta$ and
\begin{equation}
I_{g,2}(t) \approx 
\alpha_{n,2} [Q_{3} + (1 - \alpha/\gamma)Q_{2}] \exp (-\omega _{1}t),
\label{gating2b}
\end{equation}
where $\alpha_{n,2}$ is given by Eq. (\ref{alf3}).
That is, for $t \gg 1/\omega_2$, the two-stage expression for the gating current  may be 
approximated by the low frequency terms, and a similar analysis may
 also be applied to the domain of a $Na^+$ ion channel, but in order to 
account for the rising phase of the gating current that has
been observed in both \textit{Shaker} \cite{zha} and squid axon 
 $K^+$ channels \cite{wb}, as well as $Na^+$  ion channels \cite{ag}, it is necessary to take into account
the high frequency contribution.

From Eq. (\ref{rate ig}),
\begin{equation}
\frac{dI_{g,2}}{dt}(0)=\alpha [\gamma Q_{3}-(\alpha +\beta )Q_{2}],
\end{equation} 
where it is assumed that $Q_{2} \approx Q_{3}$ \cite{zha}, and for hyperpolarized
 clamp potentials, $\beta \gg \alpha ,\gamma $ and the gating
current decreases initially, whereas for depolarized potentials, $\beta$ is
small and an initial rising phase in the relaxation  of the gating current
is dependent on the relative magnitude of  the forward  rate functions 
$\alpha$  and $\gamma$. For the two-stage activation model of the 
squid axon $K^+$ ion channel, 
 $\alpha \gg \gamma$, and hence $dI_{g,2}/dt(0)<0$,  and there is no 
rising phase of the gating current for each clamp potential, whereas
for the two-stage activation model of the \textit{Shaker} $K^+$ ion channel for
which $\gamma > \alpha $, $dI_{g,2}/dt(0)>0$ and the gating current has a
rising phase for larger depolarizations \cite{zha}.

For three-stage activation of a voltage sensor, the gating charge  $Q_{g,3}(t)=Q_{2}
n_{2}(t)+(Q_{2}+Q_{3}) n_{3}(t)+(Q_{2}+Q_{3}+Q_{4}) n(t),$ and the gating current 
\begin{equation}
I_{g,3}(t)=Q_{2}\frac{dn_{2}}{dt}+(Q_{2}+Q_{3})\frac{dn_{3}}{dt}+(Q_{2}+Q_{3}+Q_{4}) \frac{dn}{dt}.
\label{gating2}
\end{equation}
For $t \gg 1/\omega_2$, the high frequency terms of the solution in Eqs. (\ref{n34})
are small  when $\omega_{3} \gg \omega_{2} \gg \omega_{1}$ (see Appendix 1) and
\begin{equation}
I_{g,3}(t) \approx \frac{\alpha_1  \exp (-\omega _{1}t)}
{(\omega_{2}-\omega_{1})(\omega_{3}-\omega_{1})}
\left[ g_1(\omega_{1})Q_{2} + (\alpha_3 + \beta_3 - \omega_{1}) \alpha_2 Q_{3} +
 \alpha_2 \alpha_3 Q_{4}\right].
\label{gating3}
\end{equation}
By taking  account of the high frequency contribution to $I_{g,3}(t)$, it may be shown from
 Eqs. (\ref{na1}) - (\ref{na}) and the initial conditions $n(0)=1$, and 
$n_{1}(0)=n_{2}(0)=n_{3}(0)=0$ that
\begin{equation}
\frac{dI_{g,3}}{dt}(0)=\alpha_1 [\alpha_2  Q_{3} - (\alpha_1 +\beta_1 )Q_{2}],
\end{equation} 
and if $\alpha_1 < \alpha_2$ and $Q_{2} \approx Q_{3}$, the gating current has a rising phase for larger depolarizations. As in the case of two-stage activation, the condition $\alpha_1 < \alpha_2$ 
 is consistent with the rate function inequality $\alpha_1 \ll \alpha_2, \alpha_3$ required for 
the rate functions (\ref{alf3st}) and (\ref{bet3st}).

\textbf{APPENDIX 3}

Stationary points of the system of Eqs. (\ref{cur1}) - (\ref{n1}) are given by
 the intersection of the  $n$, $n_{1}$ and $V$ nullclines - 
$n = (1 - n_1)\gamma/(\gamma +\delta)$, $n_1 = (1 - n)\beta/(\alpha +\beta)$ and
\begin{equation}
n = \frac{I- \overline{g}_{L}(V-V_{L})- \overline{g}_{Na}m_{\infty }(V)(V-V_{Na})}
{\overline{g}_{K}(V-V_{K})}, 
\end{equation}
and hence
\begin{equation}
\frac{I- \overline{g}_{L}(V-V_{L})- \overline{g}_{Na}m_{\infty }(V)(V-V_{Na})}
{\overline{g}_{K}(V-V_{K})}=
\frac{\alpha \gamma }{\alpha \gamma +\alpha \delta +\beta \delta },  \label{n1s}
\end{equation}
where it is assumed that the parameters are chosen so that there is only one
solution.  The stability of the stationary point may be determined by
assuming that $V=V_{s}+\widetilde{V}$, $n=n_{s}+\widetilde{n}$
and $n_{1}=n_{1s}+\widetilde{n}_{1}$, where ($V_{s},n_{s},n_{1s})$ is
the stationary solution, and defining the Jacobian matrix of coefficients $M$ of
the linearized equations in ($\widetilde{V}$, $\widetilde{n}$, $\widetilde{n}_{1})$,
\[
M = \left(
\begin{array}{ccc}
     R & -g_{K}(V_{s}-V_{K})/C & 0 \\ 
 \phi_{ns} n_{\infty s}^{\prime} (1-p_{s}) & -\phi_{ns} & 
-\phi_{ns} n_{\infty s} \\ 
   \phi_{1s}n_{1\infty s}^{\prime}(1-n_{s}) &
 -\phi_{1s} n_{1 \infty s} & -\phi_{1s}
\end{array}  \right)
\]
where $R=[-g_{L}-g_{K}n_{s}-g_{Na}m_{\infty s} - g_{Na}m_{\infty}^{\prime}(V_{s})(V_{s}-V_{Na})]/C$,
 $n_{\infty}(V)=\gamma /(\gamma +\delta )$, $n_{1\infty }(V) = \beta /(\alpha +\beta )$,
 $\phi_{ns}=\phi_{n}(V_{s})$,  $\phi_{1s} = \phi_{1}(V_{s})$, 
$n_{\infty s}= n_{\infty}(V_{s})$, $n_{1 \infty s}=n_{1 \infty}(V_{s})$, 
$n_{\infty s}^{\prime} = n_{\infty}^{\prime}(V_{s})$,
$n_{1\infty s}^{\prime} = n_{1\infty}^{\prime}(V_{s})$ and $m_{\infty s} = m_{\infty}(V_{s})$, the eigenvalues may be determined from the characteristic equation
\[
det(M-I\omega ) = \omega^{3}+f_{1} \omega^{2}+f_{2}\omega +f_{3}=0,
\]
where $I$ is the identity matrix, and
\begin{eqnarray*}
f_{1} & = & \phi_{ns}+\phi_{1s} - R, \\
f_{2} & = & -R(\phi_{ns}+\phi_{1s})-\phi_{ns}\phi_{1s}(1-n_{\infty s}n_{1\infty s})+
(g_{K}/C)(V_{s}-V_{K})\phi_{ns}n_{\infty s}^{\prime}(1-n_{1s}), \\
f_{3} & = &  (g_{K}/C)(V_{s}-V_{K})\phi_{ns} \phi_{1s}[n_{\infty s}^{\prime }(1-n_{1s}) - 
             n_{\infty s}n_{1 \infty s}^{\prime}(1-n_{s})]+ \\ 
       & &        R \phi_{ns} \phi_{1s} (n_{\infty s} n_{1 \infty s} -1).
\end{eqnarray*}

\newpage

\begin{figure*}
\begin{center}
\includegraphics[width=0.7\textwidth]{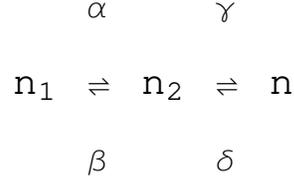}
\caption{Two-stage activation model of a $K^+$ ion channel voltage sensor, where
 the occupation probabilities of the closed states $n_{1}$, $n_{2}$
 and the open state $n$ satisfy a master equation, and  $\alpha ,\beta
, $\ $\gamma $ and  $\delta $ are voltage-dependent rate functions between
states.}
\end{center}
\end{figure*}

\begin{figure*}
\begin{center}
\includegraphics[width=0.7\textwidth]{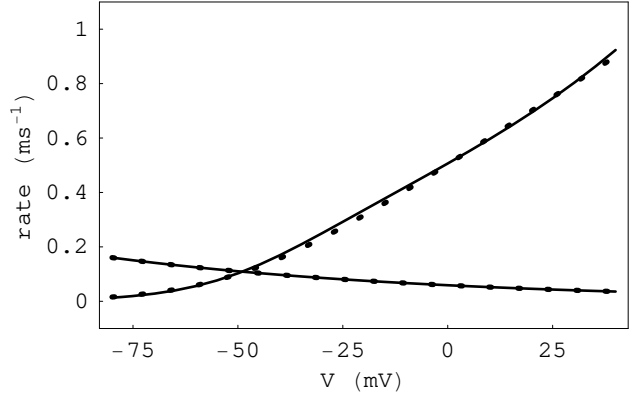}
\caption{The derived rate functions $\alpha_{n,2}$ and$\ \ \beta_{n,2}$ (solid line) in
Eqs. (\ref{alf2}) and (\ref{bet2})\ \ provide a good approximation to the HH
\ rate functions (ms$^{-1}$) $\alpha _{n}=$\ $0.01(V+50)/[1-\exp
(-0.1(V+50))]$ and $\beta _{n}=0.125\exp[-(V+60)/80]$ (dotted line) when
the rate functions for two-stage activation (ms$^{-1}$) are $\alpha
(V) = 6.4\exp [0.3(V-V_{0})/25]$, $\beta (V)=17.6\exp [-1.4(V-V_{0})/25]$, 
$\gamma (V) = 0.24 \exp$ \ $[0.345(V-V_{0})/25]$, $\delta (V) = 0.125\exp
[-0.312(V-V_{0})/25]$, $V_{0} = -57.9$.}
\end{center}
\end{figure*}

\begin{figure*}
\begin{center}
\includegraphics[width=0.7\textwidth]{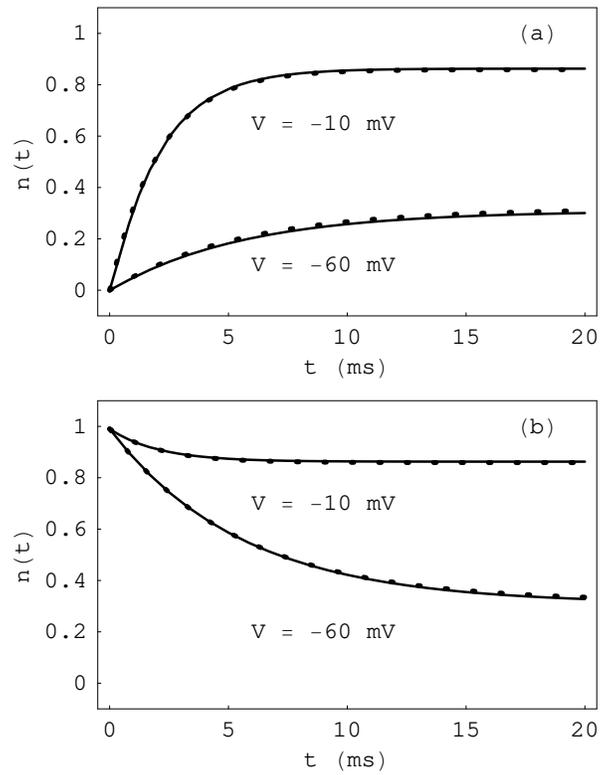}
\caption{Survival probability  of the state  $n(t)$ during a voltage clamp
 for a two-stage activation model of a $K^+$ ion channel (solid line)  
and for the HH squid axon model (dotted line)  during (a) activation
 and (b) deactivation (see Fig. 2).}
\end{center}
\end{figure*}

\begin{figure*}
\begin{center}
\includegraphics[width=0.7\textwidth]{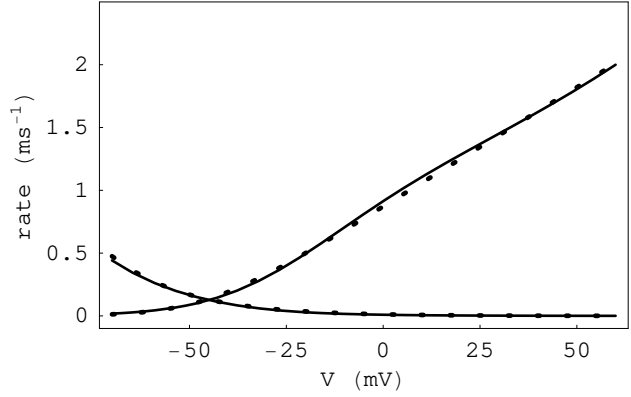}
\caption{The derived rate functions $\alpha_{n,2}$ and $\beta_{n,2}$ (solid line) in
Eqs. (\ref{alf}) and (\ref{bet}) are calculated for the rate functions
determined experimentally for the \textit{Shaker} $K^+$ ion channel \cite{zha} $
\alpha (V) = 1.1 \exp (0.25V/25)$, $\beta (V) = 0.37 \exp (-1.6V/25)$, $\gamma
(V)=2.8 \exp (0.32V/25)$, $\delta (V)=0.021 \exp (-1.1V/25)$ (ms$^{-1}$), and
may be  approximated by the rate functions $\alpha_{H}
=0.019(V+45.8)/(1-\exp [-0.15(V+45.8)])$ and  $\beta _{H}=0.135\exp
[-0.052(V+45.8)]$ for single transition activation (dotted line).}
\end{center}
\end{figure*}

\begin{figure*}
\begin{center}
\includegraphics[width=0.7\textwidth]{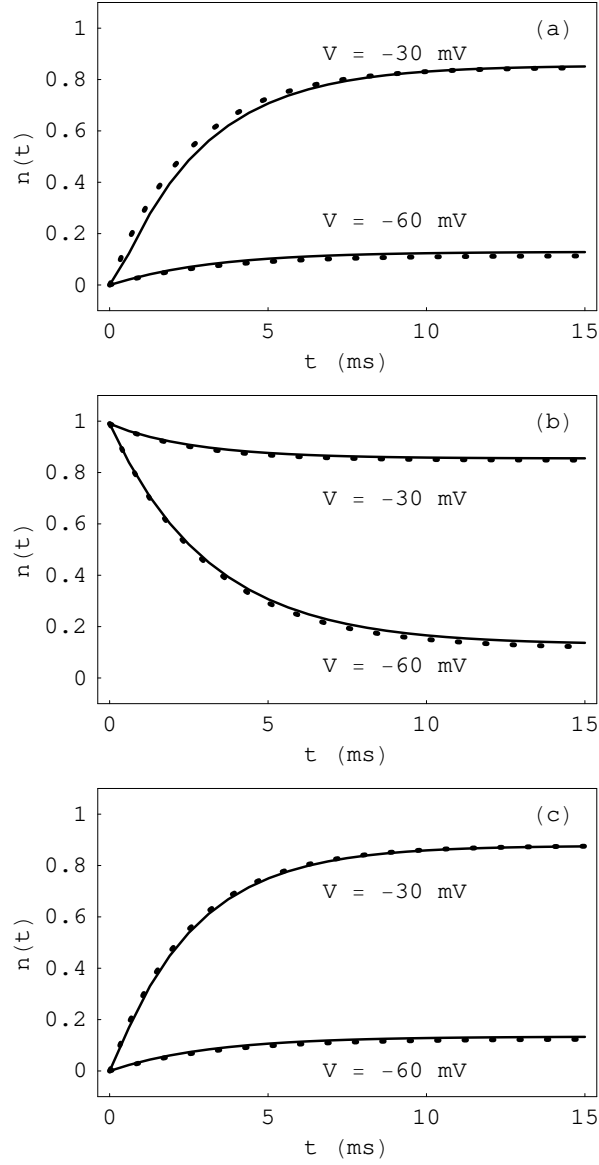}
\caption{
Survival probability  of the state $n(t)$ during a voltage clamp for a
 two-stage model of a K channel (solid line) and HH model (dotted line)
 during (a) activation and (b) deactivation, for the rate functions
 of Fig. 4,  and (c) activation when $\beta (V)$ and \ $\gamma (V)$
 are increased by a factor of 3, and the rate functions 
$\alpha_{H} = 0.02(V+45.8)/(1-\exp [-0.15(V+45.8)])$
and  $\beta_{H} = 0.12\exp [-0.057(V+45.8)].$}
\end{center}
\end{figure*}

\begin{figure*}
\begin{center}
\includegraphics[width=0.7\textwidth]{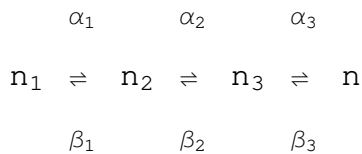}
\caption{
Three-stage activation model of a $K^+$ ion channel voltage sensor, where 
the occupation probabilities of the closed states $n_{1}$, $n_{2}$ and $n_{3}$
 and the open state $n$ satisfy a master equation, and  $\alpha_i$, $\beta_i$
for $i = 1$ to $3$  are voltage-dependent rate functions between states. }
\end{center}
\end{figure*}

\begin{figure*}
\begin{center}
\includegraphics[width=0.7\textwidth]{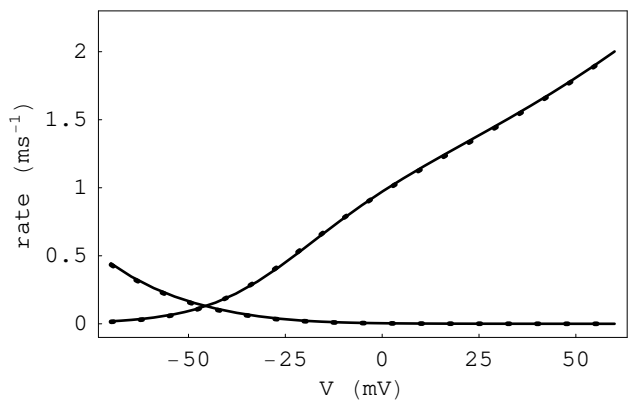}
\caption{
The derived rate functions  $\alpha_{n,2}(V)$ and $\beta_{n,2}(V)$ are calculated for 
a two-stage  $K^+$ channel (solid line), where $\alpha (V) = 1.1 \exp (0.25V/25)$,
 $\beta (V) = 1.48$\ $ \exp (-1.6V/25)$, $\gamma(V)= 11.2 \exp (0.32V/25)$, 
$\delta (V)=0.021 \exp (-1.1V/25)$ (ms$^{-1}$), and 
$\alpha_{n,3}(V)$, $\beta_{n,3}(V)$ are calculated for  a three-stage  
$K^+$ ion channel  (dotted line) where the rate functions 
$\alpha_1(V) = 1.1 \exp (0.25V/25)$, $\alpha_2(V) = 44.8 \exp$\ $(0.48$ \ $V/25)$, 
$\alpha_3(V) = 11.2 \exp (0.32V/25)$,  $\beta_1(V) = 3.0 \exp (-0.7V/25)$, 
$\beta_2(V) = 12.0 \exp (-0.7V/25)$, $\beta_3(V) = 0.021 \exp $ \ $(-1.1V/25)$.
}
\end{center}
\end{figure*}

\begin{figure*}
\begin{center}
\includegraphics[width=0.7\textwidth]{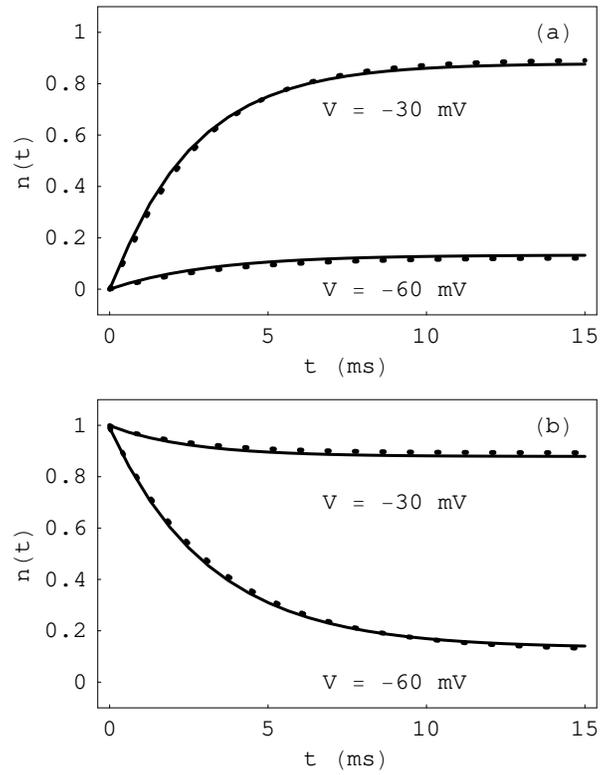}
\caption{
Survival probability  of the state $n(t)$ during a voltage clamp for a two-stage (solid line)  and three-stage (dotted line)  model of activation  of a $K^+$ ion channel voltage sensor  during (a) activation and (b) deactivation, for the rate functions of Fig. 7.
}
\end{center}
\end{figure*}

\begin{figure*}
\begin{center}
\includegraphics[width=0.8\textwidth]{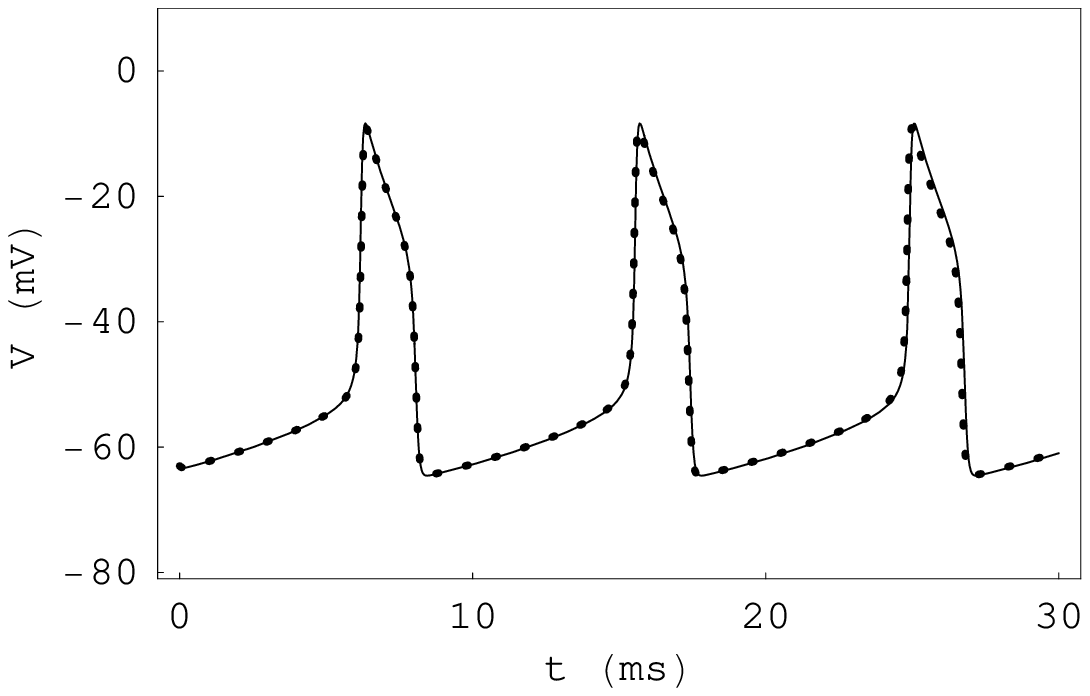}
\caption{
The limit cycle solution of Eqs. (\ref{cur1}) -  (\ref{n1}) (solid line) 
may be approximated by the solution of  Eqs. (\ref{ratek}), (\ref{alf2}), (\ref{bet2})  
 and (\ref{cur1}) (dotted line) \ when $\beta \gg \delta ,$ and $\alpha \gg \gamma .$  The rate functions $\alpha (V)=6.4\exp [0.3(V-V_{0})/25]$, 
$\beta (V)=17.6\exp [-1.4(V-V_{0})/25]$, $\gamma (V)=0.24\exp [0.345(V-V_{0})/25]$,
 $\delta (V)=0.125\exp $ \ $[-0.312(V-V_{0})/25]$, 
$\overline{g}_{K}=34$ mS/cm$^{2}$, $\overline{g}_{Na}=15$ mS/cm$^{2}$, 
$\overline{g}_{L}=0.1$ mS/cm$^{2}$, $V_{K}=-72$ mV, $V_{Na}=55$ mV, 
$V_{L}=-49.4$ mV, $I=92$ $\mu $A/cm$^{2}$, $V_{0} = -57.9$, $z_{m}=1.3$, $V_{m}=58$.
}
\end{center}
\end{figure*}

\begin{figure*}
\begin{center}
\includegraphics[width=0.8\textwidth,angle =270]{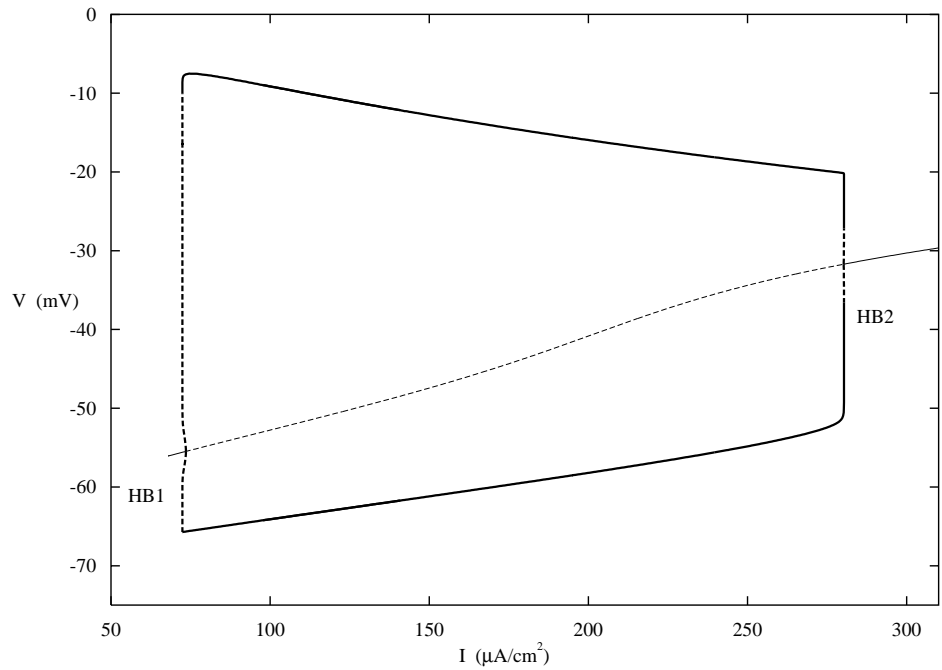}
\caption{
The bifurcation diagram for Eqs. (\ref{cur1}) - (\ref{n1}), when $\beta \gg
\delta$ and $\alpha \gg \gamma $, represents the extremal values for the stable periodic solutions (thick solid line), unstable periodic solutions (thick dashed line), and the subcritical 
bifurcation points HB1 and HB2  at the intersection between stable stationary points (thin solid line) and unstable stationary points (thin dashed line)  - see Fig. 9 for a limit cycle solution.
}
\end{center}
\end{figure*}

\begin{figure*}
\begin{center}
\includegraphics[width=0.7\textwidth]{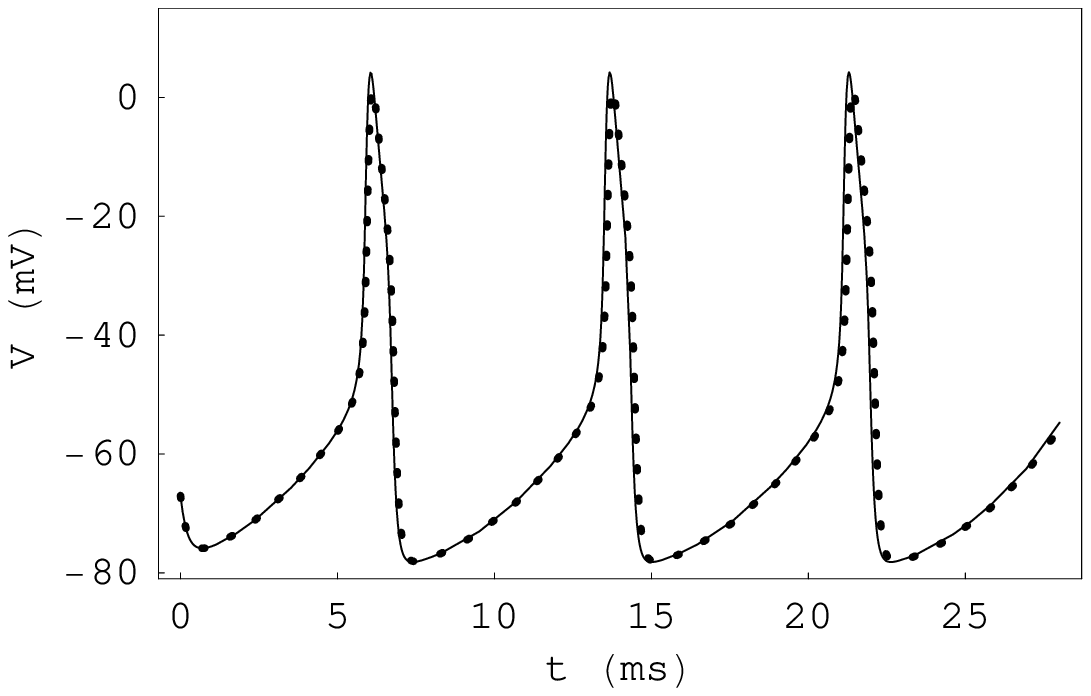}
\caption{
The limit cycle solution of Eqs. (\ref{cur1}) -\ (\ref{n1}) (solid line)
may be approximated by the solution of Eqs. (\ref{ratek}), (\ref{cur1}), (\ref{alf4})
and (\ref{bet4})  (dotted line)  when $\beta \gg \delta,$ 
and $\gamma \gg \alpha .$
 The rate functions $\alpha (V)=1.1\exp (0.25V/25)$, $\beta (V)= 1.1\exp
(-1.6V/25)$, $\gamma (V)= 8.4 \exp (0.32V/25)$, $\delta
(V)=0.021\exp (-1.1V/25)$, $\overline{g}_{K}=17$ mS/cm$^{2}$, $
\overline{g}_{Na}=9$ mS/cm$^{2}$, $\overline{g}_{L}=0.1$ mS/cm$^{2}$, $
V_{K}=-81$ mV, $V_{Na}=46$ mV, $V_{L}=-58.4$ mV, $I=13$ $\mu $A/cm$^{2}$,
 $z_{m}=1.5$, $V_{m}=51.9$.
}
\end{center}
\end{figure*}

\begin{figure*}
\begin{center}\includegraphics[width=0.8\textwidth,angle =270]{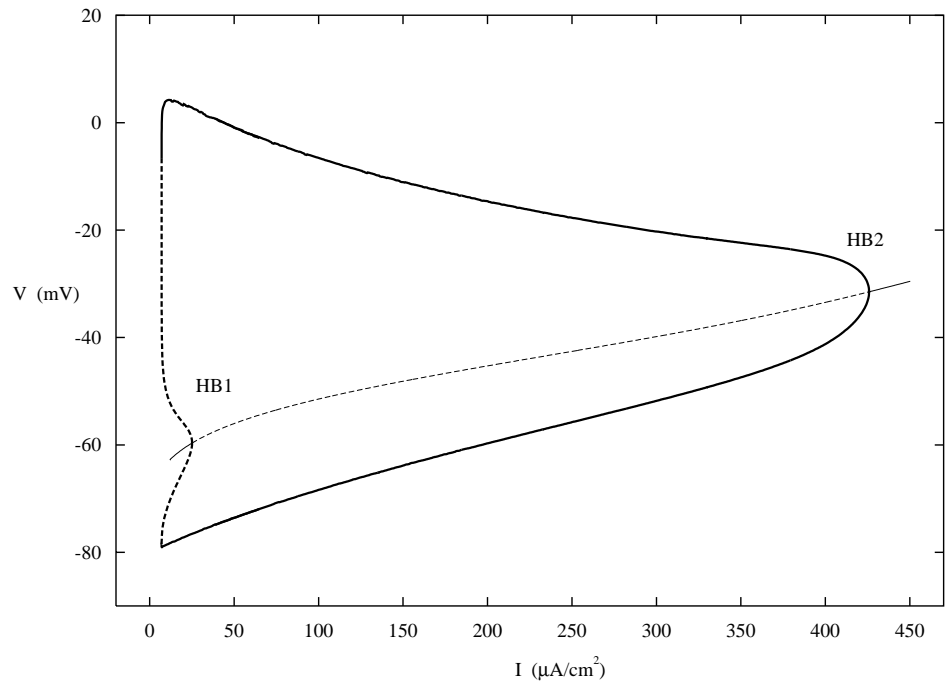}
\caption{
The bifurcation diagram for Eqs. (\ref{cur1}) -\  (\ref{n1}), when $\beta \gg
\delta$ and $\gamma \gg \alpha $,  where the subcritical (HB1) and supercritical (HB2) bifurcation points are at the junction between stable stationary points (thin solid line) and unstable stationary points (thin dashed line)  - see Fig. 11 for a limit cycle solution.
}
\end{center}
\end{figure*}

\begin{figure*}
\begin{center}
\includegraphics[width=0.8\textwidth,angle =270]{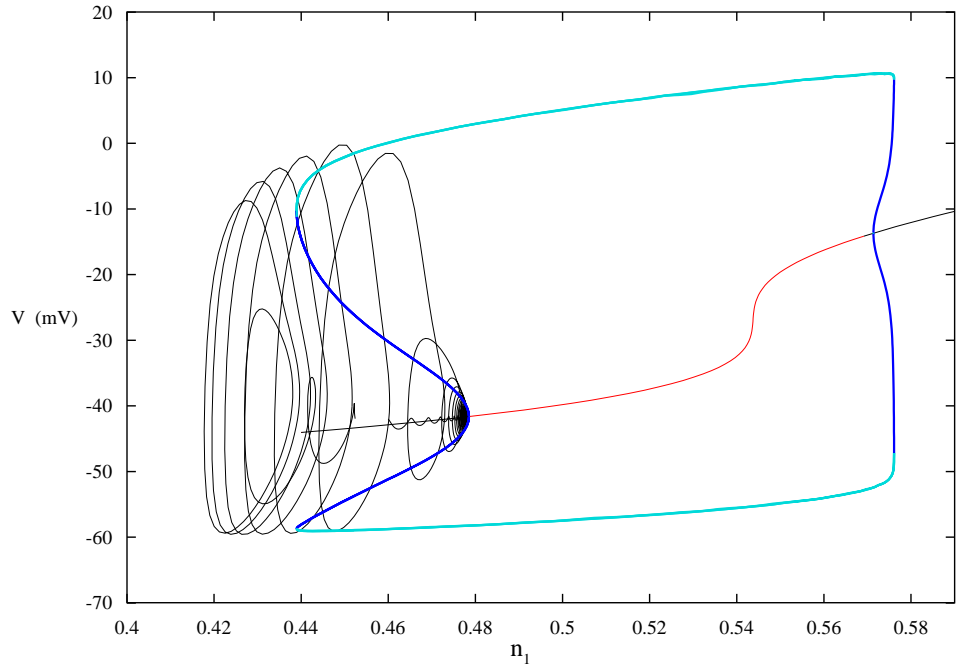}
\caption{
The bifurcation diagram for Eqs. (\ref{cur1}) and (\ref{n})  where 
$\beta \ll \delta $ and $\alpha \ll \gamma $,  the slow
variable $n_{1}$ is treated as a parameter in the (V, n) subsystem (red and
blue line), and the $V$ vs $n_{1}$\ projection is computed from 
Eqs. (\ref{cur1}) - (\ref{n1}) (solid line). For  $n_{1}$ less than $0.48$, 
 the stationary  point is stable but for  $n_{1}$ above this value, it is unstable. 
The rate functions are $\alpha (V)=0.17\exp (0.5V/25)$, 
$\beta (V)=0.02\exp (-V/25)$, $\gamma (V)=2.8\exp(0.45V/25)$,
 $\delta (V)=0.44\exp (-V/25)$, $\alpha_m(V)$ and $\beta_m(V)$ defined by Eq. (\ref{alfm}),
$V_{m}=45$, $\overline{g}_{K}=36$ mS/cm$^{2}$
, $\overline{g}_{Na}=12$ mS/cm$^{2}$, $\overline{g}_{L}=0.4$ mS/cm$^{2}$, $
V_{K}=-90$ mV, $V_{Na}=70$ mV, $V_{L}=-70$ mV, $I=236$ $\mu $A/cm$^{2}$.
}
\end{center}
\end{figure*}

\begin{figure*}
\begin{center}
\includegraphics[width=0.7\textwidth]{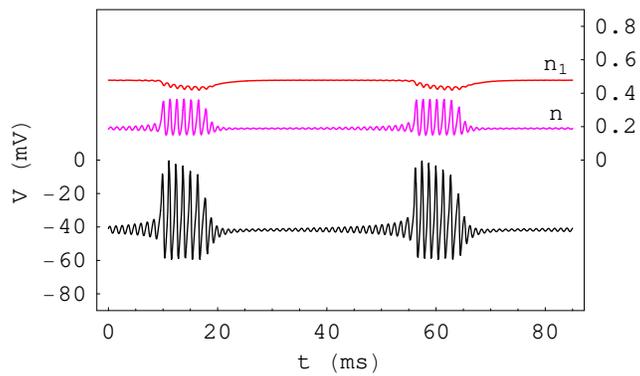}
\caption{
When $\beta \ll \delta $ and $\alpha \ll \gamma $,  Eqs. (\ref{cur1}) -  (\ref{n1})
 may exhibit a small amplitude oscillation in $V$ that alternates with a cluster 
of spikes, and is described as a bursting oscillation (see Fig. 13).
}
\end{center}
\end{figure*}

\end{document}